# Implementation of non-local XOR function for coherent-state qubit

João Batista Rosa Silva and Rubens Viana Ramos

joaobrs@deti.ufc.br          rubens@deti.ufc.br

Department of Teleinformatic Engineering, Federal University of Ceará, 60455-760, C.P. 6007, Fortaleza-Ce, Brazil

## Abstract

This work describes how to implement a non-local xor function for coherent-state qubit using only linear optics. The setup proposed does not use gates based on teleportation and it has probability of success equal to ½, in the lossless case, when the necessary entangled state is available. The key element that makes possible the realization of the non-local xor function is a tripartite GHZ-type entangled coherent state. Its generation is proposed firstly using an ideal lossless setup and secondly considering the decoherence caused by losses in the optical devices.

## 1. Introduction

Optical systems is one of the most promising technologies that can bring closer the implementation and use of quantum communication protocols and quantum computing. Even inside of optical technologies, there are different possibilities for the qubit implementation, being the single-photon polarization, phase and time-bin the most commonly qubit implementations used [1-5]. Their advantages are the facility to produce entangled states through parametric down conversion, construction of probabilistic CNOT with common optical devices and the easy implementation of single-qubit gates. The disadvantage of such qubit implementations is the fact that quantum information is carried by a single-photon, this makes the system very sensitive to losses and it requires good single-photon detectors. On the other hand the qubit implementation using superposition of coherent states has been proposed [6-9]. Such qubit implementation has as advantages the fact that it does not need single-photon detectors and the losses in the optical devices cause a quantum error (that can be corrected by a quantum code) but not the destruction of the quantum information. Its disadvantages are the hard production of coherent state superposition [10-12] and the fact that the implementation of single-qubit gates requires the teleportation procedure. In this work, we propose an optical setup for implementation of the non-local xor function for coherent state quantum information processing (CSQIP), including the optical setup for generation of the entangled state required by the protocol. In this last, two cases are considered: lossless devices (decoherence free) and lossy devices (causing decoherence).

Before starting the main work of this paper, we give a short review of CSQIP. Coherent states are eigenstates of the annihilation operator $\hat{a}$, with complex eigenvalue $\alpha$, i.e. $\hat{a}|\alpha\rangle = \alpha|\alpha\rangle$. In CSQIP, the qubit is encoded as $|0\rangle_L=|-\alpha\rangle$ and $|1\rangle_L=|\alpha\rangle$ where $\alpha$ is assumed to be real. In this case, one has $|\langle 0|1\rangle|^2=|\langle -\alpha|\alpha\rangle|^2=\exp(-4|\alpha|^2)$. Most of gates in CSQIP requires $\alpha \geq 2$, thus $|\langle \alpha|-\alpha\rangle|^2 \leq 1.11254 \times 10^{-7}$, which gives a good approximation for the orthogonality. The main optical devices used in the implementation of CSQIP are the beam splitter (BS) and the phase modulator (PM). The unitary operator of a lossless BS is $\hat{B} = \exp\left[\pi\left(\hat{a}_1\hat{a}_2^\dagger - \hat{a}_1^\dagger\hat{a}_2\right)/4\right]$. Thus, when two coherent states $|\alpha\rangle_1$ and $|\beta\rangle_2$ enter at the input ports of a balanced BS, the total state at the output ports is

$$|\alpha,\beta\rangle_{1,2} \xrightarrow{BS} \left|\frac{\alpha-\beta}{\sqrt{2}}, \frac{\alpha+\beta}{\sqrt{2}}\right\rangle_{1,2}. \qquad (1)$$

From (1), if $\beta=\alpha$ ($\beta=-\alpha$), the vacuum state appears at output mode 1(2). Hence, the setup for qubit measurement in the canonical basis consist of a BS, two common photodetectors (placed at the output ports of the BS) and a local oscillator in the state $|\alpha\rangle$. The logical state of the measured qubit is defined according to in which detector photons were received. The PM, by its turn, adds a phase $\theta$ to the signal that passes through it. Its unitary operator is $\hat{U}(\theta) = \exp(i\theta\hat{a}^\dagger\hat{a})$ and it acts like

$$|\alpha\rangle \xrightarrow{PM} |e^{j\theta}\alpha\rangle. \qquad (2)$$

Thus, if $\theta=\pi$, and the light passing by the PM is a coherent state $|\alpha\rangle$ ($-|\alpha\rangle$), then the output state will be $|-\alpha\rangle$ ($|\alpha\rangle$), thus, the PM with $\theta=\pi$ is a NOT gate in CSQIP.

The rest of this work is outlined as follows: Section 2 begins with a review of the teleportation of the xor function, after, the optical implementation of the non-local XOR function using only linear optical devices is presented; Section 3 brings the analysis of the entangled state generator when losses in the devices are considered; at last, the conclusions are presented in Section 4.

## 2. Teleportation of the xor function between two classical bits

The quantum teleportation of the xor function between two classical bits was proposed firstly in [13] and it can be used in several protocols as quantum key distribution, error correction, control of channel access and contract signature, among others. The main element of this protocol is a tripartite GHZ-type state. Initially, we consider that there are three authorized parties of the communication, Alice, Bob and Charlie, sharing the following maximally entangled tripartite of qubit state:

$$|\psi\rangle = \frac{1}{2}\left(|000\rangle_{ABC} + |011\rangle_{ABC} + |110\rangle_{ABC} + |101\rangle_{ABC}\right) \qquad (3)$$

Considering $\rho_A$, $\rho_B$ and $\rho_C$ as the individual parts of the total state $|\psi\rangle$, the teleportation of the XOR function between two classical bits, represented by $K$ (belonging to Alice) and $R$ (belonging to Bob) can be achieved using the quantum circuit shown in Fig. 1.

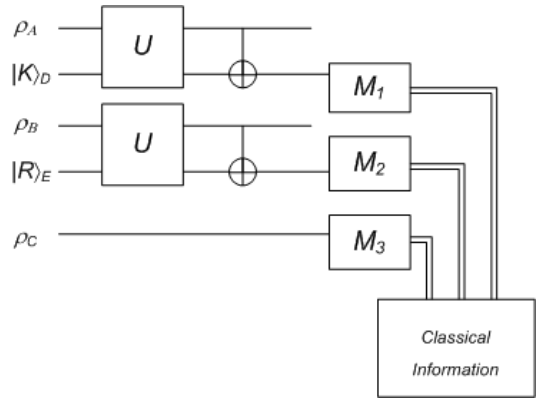

Fig. 1 - Quantum circuit for teleportation of the XOR function between two classical bits. $M_{1-3}$ are qubit measurers.

In Fig. 1, $M_1$, $M_2$ and $M_3$ are measurers while the gate $U$ represents a unitary evolution that acts in the following way:

$$U|0K\rangle = (|0K\rangle + |1\bar{K}\rangle)/\sqrt{2} \qquad (4)$$

$$U|1K\rangle = (|0\bar{K}\rangle - |1K\rangle)/\sqrt{2} \qquad (5)$$

The initial and final states are respectively given by:

$$|\Psi_{in}\rangle = |KR\rangle_{DE} \otimes \frac{1}{2}(|000\rangle_{ABC} + |011\rangle_{ABC} + |110\rangle_{ABC} + |101\rangle_{ABC}) \qquad (6)$$

$$|\Psi_{out}\rangle = \frac{1}{2}\left\{ \begin{array}{l} |++\rangle_{AB}|KR\rangle_{DE}|0\rangle_C + |+-\rangle_{AB}|K\bar{R}\rangle_{DE}|1\rangle_C + \\ |--\rangle_{AB}|\bar{K}\bar{R}\rangle_{DE}|0\rangle_C + |-+\rangle_{AB}|\bar{K}R\rangle_{DE}|1\rangle_C \end{array} \right\} \qquad (7)$$

In (7) |+⟩ and |-⟩ are simplifications of the states $(|0\rangle+|1\rangle)/2^{1/2}$ and $(|0\rangle-|1\rangle)/2^{1/2}$, respectively. When the qubits *D*, *E* and *C* are measured, by Alice, Bob and Charlie, respectively, the values {110, 101, 000, 011}$_{DEC}$ are obtained only if bits *K* and *R* are equal. On the other hand, if *K* and *R* are not equal only the values {100, 111, 010, 001}$_{DEC}$ can be obtained by the measurements. Hence, the protocol of quantum teleportation of the XOR function between two classical bits can be described as follows:

- Alice performs a measurement in the qubit *D* and she sends her result to Charlie using one classical bit.
- Bob performs a measurement in the qubit *E* and he sends his result to Charlie using another classical bit.
- Charlie, by its turn, performs a measurement in his qubit. Knowing those three classical information, Charlie can know if *K* and *R* are equal or not. Hence, the XOR function between the classical bits belonging to Alice and Bob is teleported to Charlie.

The classical bits sent by Alice and Bob inform to Charlie not the values of *K* and *R*, but if *K* and *R* are equal or not to the individual states $\rho_A$ and $\rho_B$, respectively.

As seen before, in order to realize the teleportation of the xor function between two classical bits, the first task to be done is the generation of the tripartite entangled state (3). In CSQIP, the state (3) is rewritten as

$$|\psi\rangle = \frac{1}{2}(|-\alpha,-\alpha,-\alpha\rangle + |-\alpha,\alpha,\alpha\rangle + |\alpha,-\alpha,\alpha\rangle + |\alpha,\alpha,-\alpha\rangle). \qquad (8)$$

The state (8) can be obtained by application of a Hadamard gate in each individual state of the tripartite GHZ state $(|-\alpha,-\alpha,-\alpha\rangle+|\alpha,\alpha,\alpha\rangle)/2^{1/2}$. This last one can be generated by the optical setup proposed in [6,14] and shown in Fig. 2, where the beam splitters BS1 and BS2 have, respectively, reflectivity equal to $3^{-1/2}$ and $2^{-1/2}$.

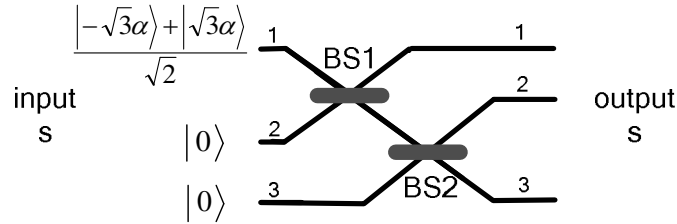

Fig. 2 – Entanglement generator circuit for GHZ state $(|-\alpha,-\alpha,-\alpha\rangle+|\alpha,\alpha,\alpha\rangle)/2^{1/2}$.

By its turn, the Hadamard gate in coherent state qubit has to realize the transformations $|-\alpha\rangle \rightarrow (|-\alpha\rangle+|\alpha\rangle)/2^{1/2}$ and $|\alpha\rangle \rightarrow (|-\alpha\rangle-|\alpha\rangle)/2^{1/2}$. This operation takes non-orthogonal states to orthogonal states and, hence, it is not unitary. Basically, the Hadamard gate needs a teleportation that, sometimes, requires a Z gate that, by its turn, is implemented realizing a teleportation. Hence, the generation of the quantum state (8) using the setup in Fig. 2 and three Hadamard gates has a low efficiency. Due to the low efficiency this strategy, we propose the setup shown in Fig. 3 for generation of the state in (8).

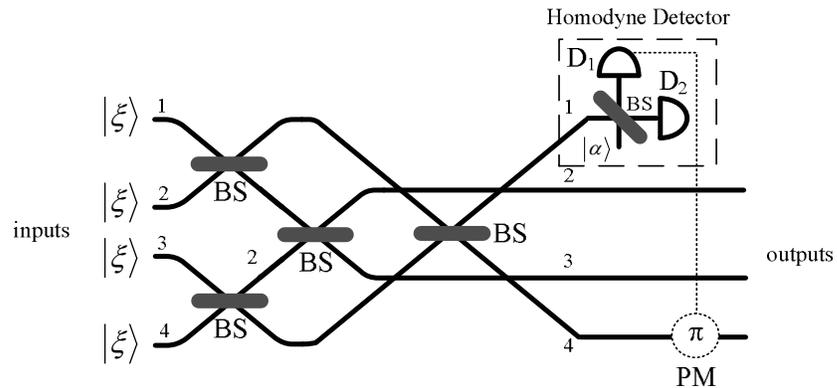

Fig. 3 – Optical scheme for generation of the tripartite entangled state given in (8).

In Fig. 3, $|\xi\rangle = N(|-\alpha\rangle + |\alpha\rangle)/2^{1/2}$ where $N=[2(1+e^{-2\alpha^2})]^{-1/2}$ is the normalization constant and BS are balanced beam splitter. After some trivial calculations, one can find the following output state before the homodyne detection

$$|\Psi_o\rangle = \frac{1}{\sqrt{2}}\left[|\alpha\rangle_1|\psi_1\rangle + |-\alpha\rangle_1|\psi_2\rangle\right] + \frac{1}{\sqrt{2}}|\Psi_u\rangle \tag{9}$$

$$|\psi_1\rangle = \frac{1}{2}(|-\alpha,-\alpha,-\alpha\rangle + |-\alpha,\alpha,\alpha\rangle + |\alpha,-\alpha,\alpha\rangle + |\alpha,\alpha,-\alpha\rangle)_{234} = |\psi\rangle \tag{10}$$

$$|\psi_2\rangle = (I \otimes I \otimes X)|\psi\rangle \tag{11}$$

In (9) $|\Psi_u\rangle$ is the useless part that contains the situations where detection happen in both detectors, $D_1$ and $D_2$, in this case, the circuit fails. From (7) one can also note that when the homodyne detector measures $|\alpha\rangle$, the output is $|\psi\rangle$ and the X gate (PM) is disabled. On the other hand, if the result of the measurement is $|-\alpha\rangle$, then the X gate is activated in order to correct the output state, according to (11). The probability of success of the setup shown in Fig. 3 is 1/2.

Once the required tripartite state was generated, one can use the setup in Fig. 4 to implement the quantum circuit presented in Fig. 1 and run the non-local xor function protocol.

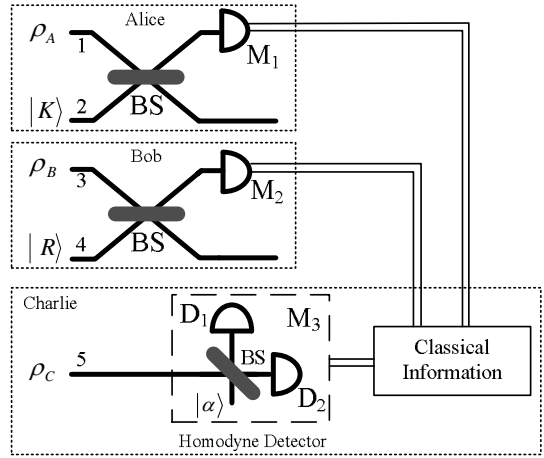

Fig. 4 – Optical setup for teleportation of the xor function of two classical bits.

The input state is $[2^{-1}(|-\alpha,-\alpha,\alpha\rangle + |-\alpha,\alpha,\alpha\rangle + |\alpha,-\alpha,\alpha\rangle + |\alpha,\alpha,-\alpha\rangle)_{135}] \otimes |K\rangle_2 \otimes |R\rangle_4$, where K and R ∈ $\{-\alpha,\alpha\}$. The total quantum state just before the measurements is

$$|\Psi_o\rangle = \frac{1}{2}\left(\left|\frac{-(\alpha+K)}{\sqrt{2}},\frac{K-\alpha}{\sqrt{2}},\frac{-(\alpha+R)}{\sqrt{2}},\frac{R-\alpha}{\sqrt{2}},-\alpha\right\rangle + \left|\frac{-(\alpha+K)}{\sqrt{2}},\frac{K-\alpha}{\sqrt{2}},\frac{\alpha-R}{\sqrt{2}},\frac{R+\alpha}{\sqrt{2}},\alpha\right\rangle + \left|\frac{\alpha-K}{\sqrt{2}},\frac{K+\alpha}{\sqrt{2}},\frac{-(\alpha+R)}{\sqrt{2}},\frac{R-\alpha}{\sqrt{2}},\alpha\right\rangle + \left|\frac{\alpha-K}{\sqrt{2}},\frac{K+\alpha}{\sqrt{2}},\frac{\alpha-R}{\sqrt{2}},\frac{R+\alpha}{\sqrt{2}},-\alpha\right\rangle\right)_{12345} \quad (12)$$

The measurers $M_1$ and $M_2$ just detect the presence or absence of the light meaning, respectively, bit '1' and bit '0'. Charlie uses a measurer based on homodyne detection. Therefore, when the qubits *A*, *B* and *C* are measured, by Alice, Bob and Charlie, respectively, the values {110, 101, 000, 011}$_{ABC}$ are obtained only if qubits *K* and *R* are equal. On the other hand, if *K* and *R* are not equal only the values {100, 111, 010, 001}$_{ABC}$ can be obtained by the measurement conform described by the quantum teleportation protocol of the xor function between two classical bits.

## 3. Entanglement generation under decoherence caused by lossy devices

It has been shown that losses cause decoherence in CSQIP [6]. In order to have an idea of the effect of the losses in the generation of the tripartite entangled state given in (8) we consider the optical scheme presented in Fig. 3 with lossy beam splitters. In order to do this, we follow the stratagem used in [6] that models the loss by a lossless beam splitter with vacuum state at one of the input ports. Thus, the new optical setup to be considered is the one shown in Fig. 5.

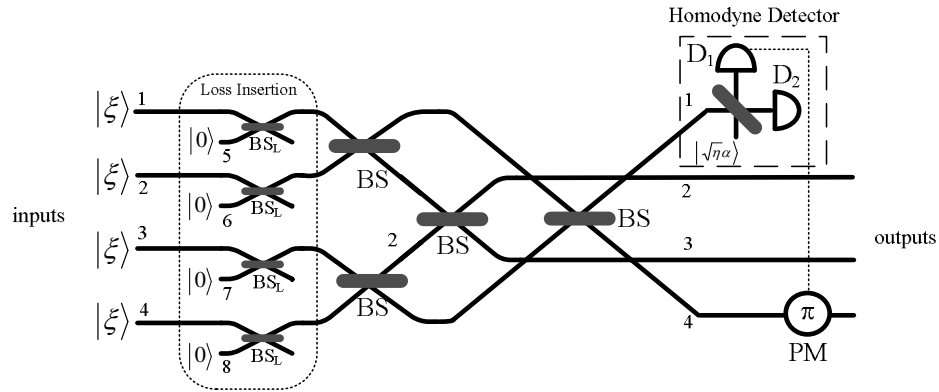

Fig. 5 – Optical setup for entanglement generation using lossy devices, where the losses are modeled by ideal beam splitters at the inputs.

In Fig. 5, the ideal beam splitters $BS_L$ are placed at the inputs in order to model the losses in the other optical devices of the setup. Without loss of generality, all $BS_L$ are considered to have the same transmissivity, $\eta$.

After some algebra, the total state before homodyne detection in the optical system in Fig. 5 is given by

$$|\Psi_{oL}\rangle = \frac{1}{\sqrt{2}}\left[|\sqrt{\eta}\alpha\rangle_1|\psi_{1L}\rangle + |-\sqrt{\eta}\alpha\rangle_1|\psi_{2L}\rangle\right] + \frac{1}{\sqrt{2}}|\Psi_{uL}\rangle \quad (13)$$

$$|\psi_{1L}\rangle = 2N^4 \begin{pmatrix} |-\sqrt{\eta}\alpha,-\sqrt{\eta}\alpha,-\sqrt{\eta}\alpha\rangle_{234}|-\sqrt{1-\eta}\alpha,-\sqrt{1-\eta}\alpha,-\sqrt{1-\eta}\alpha,-\sqrt{1-\eta}\alpha\rangle_{5678} + \\ |-\sqrt{\eta}\alpha,\sqrt{\eta}\alpha,\sqrt{\eta}\alpha\rangle_{234}|\sqrt{1-\eta}\alpha,-\sqrt{1-\eta}\alpha,\sqrt{1-\eta}\alpha,-\sqrt{1-\eta}\alpha\rangle_{5678} + \\ |\sqrt{\eta}\alpha,-\sqrt{\eta}\alpha,\sqrt{\eta}\alpha\rangle_{234}|\sqrt{1-\eta}\alpha,-\sqrt{1-\eta}\alpha,-\sqrt{1-\eta}\alpha,\sqrt{1-\eta}\alpha\rangle_{5678} + \\ |\sqrt{\eta}\alpha,\sqrt{\eta}\alpha,-\sqrt{\eta}\alpha\rangle_{234}|\sqrt{1-\eta}\alpha,\sqrt{1-\eta}\alpha,-\sqrt{1-\eta}\alpha,-\sqrt{1-\eta}\alpha\rangle_{5678} \end{pmatrix} \quad (14)$$

$$|\psi_{2L}\rangle = 2N^4 \begin{pmatrix} |-\sqrt{\eta}\alpha,-\sqrt{\eta}\alpha,\sqrt{\eta}\alpha\rangle_{234}|-\sqrt{1-\eta}\alpha,-\sqrt{1-\eta}\alpha,\sqrt{1-\eta}\alpha,\sqrt{1-\eta}\alpha\rangle_{5678} + \\ |-\sqrt{\eta}\alpha,\sqrt{\eta}\alpha,-\sqrt{\eta}\alpha\rangle_{234}|-\sqrt{1-\eta}\alpha,\sqrt{1-\eta}\alpha,\sqrt{1-\eta}\alpha,-\sqrt{1-\eta}\alpha\rangle_{5678} + \\ |\sqrt{\eta}\alpha,-\sqrt{\eta}\alpha,-\sqrt{\eta}\alpha\rangle_{234}|-\sqrt{1-\eta}\alpha,\sqrt{1-\eta}\alpha,-\sqrt{1-\eta}\alpha,\sqrt{1-\eta}\alpha\rangle_{5678} + \\ |\sqrt{\eta}\alpha,\sqrt{\eta}\alpha,\sqrt{\eta}\alpha\rangle_{234}|\sqrt{1-\eta}\alpha,\sqrt{1-\eta}\alpha,\sqrt{1-\eta}\alpha,\sqrt{1-\eta}\alpha\rangle_{5678} \end{pmatrix}_{234} \Rightarrow \quad (15)$$

$$|\psi_{2L}\rangle = (I \otimes I \otimes X)_{234} \otimes (I \otimes I \otimes I \otimes I)_{5678}|\psi_{1L}\rangle. \quad (16)$$

The state $|\Psi_{uL}\rangle$ contains the situations where detections occur in both detectors, meaning that a vacuum state is present in at least one of the output ports. Now, using $\alpha \geq 2\eta^{-1/2}$ one can define the new logical states $|0\rangle_L \equiv |-\eta^{1/2}\alpha\rangle$ and $|1\rangle_L \equiv |\eta^{1/2}\alpha\rangle$. Furthermore, using $\beta^{\pm} = \pm(1-\eta)^{1/2}\alpha$, one can rewrite (13)-(15) as

$$|\Psi_{oL}\rangle = \frac{1}{\sqrt{2}}[|1\rangle_1|\psi_{1L}\rangle + |0\rangle_1|\psi_{2L}\rangle] + \frac{1}{\sqrt{2}}|\Psi_{uL}\rangle \quad (17) \quad\quad (5)$$

$$|\psi_{1L}\rangle = 2N^4(|000\rangle_{234}|\beta^-\beta^-\beta^-\beta^-\rangle_{5678} + |011\rangle_{234}|\beta^+\beta^-\beta^+\beta^-\rangle_{5678} + |101\rangle_{234}|\beta^+\beta^-\beta^-\beta^+\rangle_{5678} + |110\rangle_{234}|\beta^+\beta^+\beta^-\beta^-\rangle_{5678}) \quad (18) \quad (6)$$

$$|\psi_{2L}\rangle = 2N^4(|001\rangle_{234}|\beta^-\beta^-\beta^+\beta^+\rangle_{5678} + |010\rangle_{234}|\beta^-\beta^+\beta^+\beta^-\rangle_{5678} + |100\rangle_{234}|\beta^-\beta^+\beta^-\beta^+\rangle_{5678} + |111\rangle_{234}|\beta^+\beta^+\beta^+\beta^+\rangle_{5678}) \quad (19) \quad (7)$$

Without taking into account the state $|\Psi_{uL}\rangle$, which is rejected after the measurements, the valid output state, which happens with probability ½, is

$$|\psi'_{oL}\rangle = 2^{-1/2}[|1\rangle_1|\psi_{1L}\rangle + |0\rangle_1|\psi_{2L}\rangle], \quad \rho_{oL} = |\psi'_{oL}\rangle\langle\psi'_{oL}| = \frac{1}{2}\begin{bmatrix}|\psi_{1L}\rangle|1\rangle_1\langle 1|_1\langle\psi_{1L}| + |\psi_{1L}\rangle|1\rangle_1\langle 0|_1\langle\psi_{2L}| + \\ |\psi_{2L}\rangle|0\rangle_1\langle 1|_1\langle\psi_{1L}| + |\psi_{2L}\rangle|0\rangle_1\langle 0|_1\langle\psi_{2L}|\end{bmatrix}. \quad (20)$$

Now, tracing out the "lost" modes (qubits 5-8), the useful tripartite output state is

$$\rho_o = Tr_{5678}(\rho_{oL}) = \frac{1}{2}\begin{bmatrix} Tr_{5678}(|\psi_{1L}\rangle|1\rangle_1\langle 1|_1\langle\psi_{1L}|) + Tr_{5678}(|\psi_{1L}\rangle|1\rangle_1\langle 0|_1\langle\psi_{2L}|) + \\ Tr_{5678}(|\psi_{2L}\rangle|0\rangle_1\langle 1|_1\langle\psi_{1L}|) + Tr_{5678}(|\psi_{2L}\rangle|0\rangle_1\langle 0|_1\langle\psi_{2L}|) \end{bmatrix}. \quad (21)$$

Naming $|\psi\rangle = 2N^4(|-\alpha,-\alpha,-\alpha\rangle + |-\alpha,\alpha,\alpha\rangle + |\alpha,-\alpha,\alpha\rangle + |\alpha,\alpha,-\alpha\rangle)_{234}$, $|\psi'\rangle = (I \otimes I \otimes X)|\psi\rangle$ and having $\langle \beta^\pm | \beta^\pm \rangle = 1$ and $\langle \beta^\pm | \beta^\mu \rangle = \delta^{1/2}$ where $\delta = \exp[-4\alpha^2(1-\eta)]$, one can find

$$Tr_{5678}(|\psi_{1L}\rangle|1\rangle_1\langle 1|_1\langle\psi_{1L}|) = 4N^8 \begin{bmatrix} (|1000\rangle\langle 1000| + |1011\rangle\langle 1011| + |1101\rangle\langle 1101| + |1110\rangle\langle 1110|) + \\ \delta \begin{pmatrix} |1000\rangle\langle 1101| + |1000\rangle\langle 1011| + |1000\rangle\langle 1110| + \\ |1101\rangle\langle 1000| + |1101\rangle\langle 1011| + |1101\rangle\langle 1110| + \\ |1011\rangle\langle 1000| + |1011\rangle\langle 1101| + |1011\rangle\langle 1110| + \\ |1110\rangle\langle 1000| + |1110\rangle\langle 1101| + |1110\rangle\langle 1011| \end{pmatrix} \end{bmatrix}$$
$$= \delta \cdot (|\psi_1\rangle|1\rangle_1\langle 1|_1\langle\psi_1|) + (1-\delta)\cdot \rho_1, \quad (22)$$

$$Tr_{5678}(|\psi_{1L}\rangle|1\rangle_1\langle 0|_1\langle\psi_{2L}|) = 4N^8 \begin{bmatrix} \delta^2(|1000\rangle\langle 0111| + |1101\rangle\langle 0010| + |1011\rangle\langle 0100| + |1110\rangle\langle 0001|) + \\ \delta \begin{pmatrix} |1000\rangle\langle 0001| + |1000\rangle\langle 0100| + |1000\rangle\langle 0010| + \\ |1101\rangle\langle 0001| + |1101\rangle\langle 0100| + |1101\rangle\langle 0111| + \\ |1011\rangle\langle 0001| + |1011\rangle\langle 0010| + |1011\rangle\langle 0111| + \\ |1110\rangle\langle 0100| + |1110\rangle\langle 0010| + |1110\rangle\langle 0111| \end{pmatrix} \end{bmatrix}$$
$$\cong 4N^8 \delta(|\psi_1\rangle|1\rangle_1\langle 0|_1\langle\psi_2|), \quad (23)$$

$$Tr_{5678}(|\psi_{2L}\rangle|0\rangle_1\langle 1|_1\langle\psi_{1L}|) = Tr_{5678}(|\psi_{1L}\rangle|1\rangle_1\langle 0|_1\langle\psi_{2L}|)^\dagger \cong 4N^8 \delta(|\psi_2\rangle|0\rangle_1\langle 1|_1\langle\psi_1|) \quad (24)$$

$$Tr_{5678}(|\psi_{2L}\rangle|0\rangle_1\langle 0|_1\langle\psi_{2L}|) = 4N^8 \begin{bmatrix} (|0001\rangle\langle 0001| + |0100\rangle\langle 0100| + |0010\rangle\langle 0010| + |0111\rangle\langle 0111|) + \\ \delta \begin{pmatrix} |0001\rangle\langle 0100| + |0001\rangle\langle 0010| + |0001\rangle\langle 0111| + \\ |0100\rangle\langle 0001| + |0100\rangle\langle 0010| + |0100\rangle\langle 0111| + \\ |0010\rangle\langle 0001| + |0010\rangle\langle 0100| + |0010\rangle\langle 0111| + \\ |0111\rangle\langle 1000| + |0111\rangle\langle 0100| + |0111\rangle\langle 0010| \end{pmatrix} \end{bmatrix}$$
$$= \delta \cdot (|\psi_2\rangle|0\rangle_1\langle 0|_1\langle\psi_2|) + (1-\delta)\cdot \rho_2. \quad (25)$$

We have used $\delta^2 \approx \delta$ in (23) and (24) and

$$\rho_1 = 4N^8 \cdot (|1000\rangle\langle 1000| + |1011\rangle\langle 1011| + |1101\rangle\langle 1101| + |1110\rangle\langle 1110|) \quad (26)$$

$$\rho_2 = 4N^8 \cdot (|0001\rangle\langle 0001| + |0100\rangle\langle 0100| + |0010\rangle\langle 0010| + |0111\rangle\langle 0111|). \quad (27)$$

Finally, (21) can be written as

$$\rho_o = \frac{1}{2}\{\delta[|\psi\rangle|\sqrt{\eta}\alpha\rangle_1\langle\sqrt{\eta}\alpha|_1\langle\psi| + U|\psi\rangle|-\sqrt{\eta}\alpha\rangle_1\langle-\sqrt{\eta}\alpha|_1\langle\psi|U^\dagger] + (1-\delta)(\rho_1 + \rho_2)\}, \qquad (28)$$

where $U = I \otimes I \otimes X$ and the total probability of success is $\delta/2$.

In Fig. 6, one can observe the relation between the probability of success $\delta/2$ and the transmissivity $\eta$ of the beam splitters that model the losses, for three different values of $\alpha$.

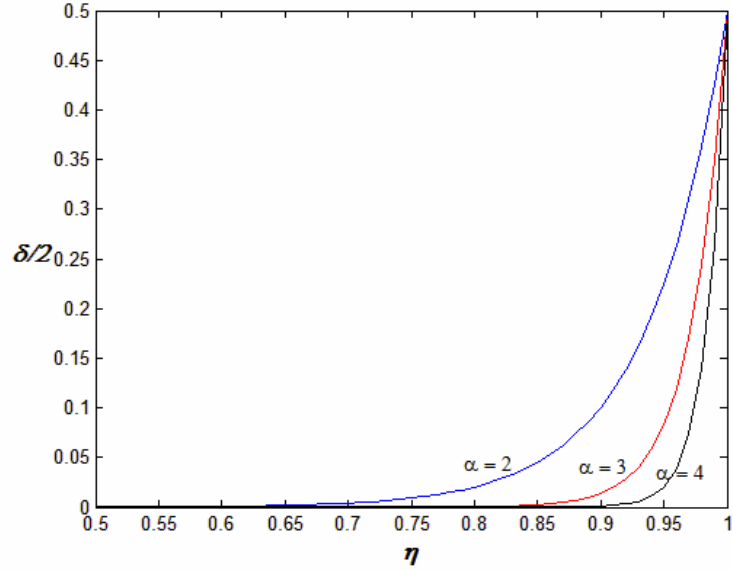

Fig. 6 – Probability of success versus beam splitter transmissivity.

In [6] it was shown that a (lossy) Hadamard gate based on teleportation succeeds with probability 0.29 (0.59) for $\eta=0.8536$ ($\eta=0.9904$) and $\alpha=2$ ($\alpha=4$). In this case, if one tries to obtain the state (8) using the (lossless) setup in Fig. 2 and three (lossy) Hadarmard gates, the total probability of success is 0.0244 (0.2054). On the other hand, using the setup shown in Fig. 5 with $\alpha=2$ (4) and $\eta=0.8536$ (0.9904), the probability of success is 0.0475 (0.2636).

At last, if one places the beam splitters $BS_L$ at the outputs instead of the inputs, the same results are obtained.

## 5. Conclusions

Firstly, we proposed a setup for probabilistic generation of the tripartite state (8) that it is used for the realization of the quantum teleportation protocol of the xor function between two classical bits. In the sequence, we presented a proposal of realization of the quantum teleportation protocol of the xor function, for coherent state qubit, using only linear optical devices. The efficiency of the proposed (lossless) teleporter setup is 1/2 if one considers that the used entangled state (8) is available. An advantage of our setups is the absence of single-qubit gates based on teleportation, that are common in quantum information processing with coherent states. At last, it was analyzed the decoherence effects caused by lossy devices in the entangled state generator. It was observed that, in order to guarantee a probability of success upper than 0.25, the transmissivity $\eta$ should not be lower than 0.96, implying that even low losses can be harmful for the efficiency of the setup.

## Acknowledgements

Useful discussions with Hilma Vasconcelos are gratefully acknowledged.